\documentclass[12pt,preprint,natbib209]{aastex}
\slugcomment{To appear in ApJ}

\shorttitle{X-RAYS FROM THERMONUCLEAR SNRs II}
\shortauthors{Badenes, Borkowski and Bravo}

\begin{document}

\title{Thermal X-Ray Emission from Shocked Ejecta in Type Ia Supernova Remnants II:\\ Parameters Affecting the Spectrum}

\author{Carles Badenes\altaffilmark{1}, Kazimierz J. Borkowski\altaffilmark{2} and Eduardo Bravo\altaffilmark{3,4}}

\altaffiltext{1}{Department of Physics and Astronomy, Rutgers University, 136 Frelinghuysen Rd., 
  Piscataway NJ 08854-8019; badenes@physics.rutgers.edu}

\altaffiltext{2}{Department of Physics, North Carolina State University, Box 8202, Raleigh NC 27965-8202; kborkow@unity.ncsu.edu}

\altaffiltext{3}{Departament de F\'{i}sica i Enginyeria Nuclear, Universitat Polit\`{e}cnica de Catalunya, Diagonal 647, 
  Barcelona 08028, Spain; eduardo.bravo@upc.es}

\altaffiltext{4}{Institut d'Estudis Espacials de Catalunya, Campus UAB, Facultat de Ci\`{e}ncies. Torre C5. Bellaterra,  
  Barcelona 08193, Spain}

\begin{abstract}
The supernova remnants left behind by Type Ia supernovae provide an excellent opportunity for the study of these enigmatic
objects. In a previous work, we showed that it is possible to use the X-ray spectra of young Type Ia supernova
remnants to explore the physics of Type Ia supernovae and identify the relevant mechanism underlying these explosions. 
Our simulation technique is based on hydrodynamic and nonequilibrium ionization calculations of the interaction of a grid of 
Type Ia explosion models with the surrounding ambient medium, coupled to an X-ray spectral code. In this work we explore the 
influence of two key parameters on the shape of the X-ray spectrum of the ejecta: the density of the ambient medium around the 
supernova progenitor and the efficiency of collisionless electron heating at the reverse shock. We also discuss the performance
of recent 3D simulations of Type Ia SN explosions in the context of the X-ray spectra of young SNRs. We find a better agreement 
with the observations for Type Ia supernova models with stratified ejecta than for 3D deflagration models with well mixed ejecta. 
We conclude that our grid of Type Ia supernova remnant models can improve our understanding of these objects and their relationship 
to the supernovae that originated them. 

\end{abstract}


\section{INTRODUCTION}

The advent of modern X-ray observatories such as {\it Chandra} and {\it XMM-Newton} has produced a spectacular 
increase in both the quantity and the quality of the observations of Type Ia supernova remnants (SNRs). Yet, these
excellent observations have led only to a modest improvement in our knowledge of the physics of Type Ia supernovae (SNe).
Important issues such as the nature of the progenitor systems, the last stages of their evolution prior to the SN 
explosion or the physical mechanism behind the explosion itself still remain obscure 
\citep[see][for reviews]{hillebrandt00:Ia-review,branch95:Ia-review}. In a recent paper \citep[][henceforth Paper I]{badenes03:xray}, 
we examined the prospects for the identification of the explosion mechanism in Type Ia SNe through the analysis of the X-ray spectra 
of young SNRs. We assembled a grid of Type Ia SN explosion models, simulated their interaction with a uniform 
ambient medium (AM) and calculated the predicted X-ray spectra from the ensuing SNRs. The calculated X-ray SNR spectra varied
dramatically from model to model, demonstrating that it is possible to use young Type Ia SNRs to probe the details of the Type Ia SN 
explosion mechanism. 

In this paper, we expand the results that were introduced in Paper I. Our objective is to
examine the relationship between Type Ia SN explosions and the X-ray spectra of their SNRs within
the framework of hydrodynamic, ionization, and spectral simulations. By comparing our models with observations, we aim at 
improving our understanding of both Type Ia SNe and young, ejecta-dominated SNRs. In \S~\ref{sec:Techniques}, we review 
the simulation scheme used in Paper I, and we discuss the influence of two important parameters which we had not hitherto explored: 
the amount of collisionless electron heating at the reverse shock and the density of the uniform  AM that interacts with the 
ejecta. In \S~\ref{sec:Expanding} we examine the performance of recent 3D Type Ia SN explosion models in the context
of SNRs, and we discuss the ability of these 3D models to reproduce the fundamental properties of 
the X-ray spectra of Type Ia SNRs. Our conclusions are presented in \S~\ref{sec:Discussion}. In order to facilitate the 
comparison between our models and X-ray observations of SNRs, we have generated a library of synthetic spectra. This library is 
presented and discussed in the Appendix. In a forthcoming paper (Badenes et al., in preparation), we will make a detailed 
comparison between our models and the X-ray spectrum of the Tycho SNR.

\section{PARAMETERS AFFECTING THE X-RAY SPECTRUM} \label{sec:Techniques}

\subsection{From SN to SNR: the Simulation Scheme} 

Although the X-ray spectra of young Type Ia SNRs contain much information about the structure and composition of 
the material ejected by the SNe that originated them, this information is generally difficult to extract. The ejecta material 
consists almost entirely of heavy elements that are impulsively heated to X-ray emitting temperatures as the reverse shock 
propagates inwards in the reference frame of the expanding ejecta. The propagation of the reverse shock is in turn 
intimately related to the density structure of the ejecta, which results in an intricate dynamical behavior of the SNR early
in its evolution \citep[see][]{dwarkadas98:typeIa}. If there is a significant degree of stratification 
in the elemental composition of the ejecta, different chemical elements are shocked at different evolutionary times, after 
different periods of free expansion, and therefore emit X-rays under different physical conditions. This
results in a very complex spectrum, which is hard to model and interpret.

The approach taken in Paper I was based on a grid of 1D Type Ia SN explosion models that included all the mechanisms currently 
under debate for the single degenerate Type Ia SN scenario: deflagrations, delayed detonations, pulsating delayed detonations and 
sub-Chandrasekhar explosions. The dynamics of the interaction of each explosion model with a uniform 
ambient medium (AM) of density $\rho_{AM}=10^{-24}\,\mathrm{g\cdot cm^{-3}}$ was followed with a 1D hydrodynamic code. 
The dynamic evolution of each fluid element in the shocked ejecta (i.e., the time evolution of density $\rho$ and specific 
internal energy per unit mass $\varepsilon$), together with its chemical composition as determined by the SN explosion model,
were used as the input to ionization calculations. These calculations included the interactions between ions and electrons in the 
shocked plasma, and they provided time-dependent nonequilibrium ionization (NEI) states and electron temperature. Using these 
ionization states and electron temperatures, spatially integrated synthetic X-ray spectra were generated with a spectral code 
by adding the weighted contributions from each fluid element in the shocked ejecta. For a more detailed explanation, see Paper I 
and the references therein. 

Within this simulation scheme, the X-ray spectrum from the shocked ejecta is determined by: (1) the density 
and chemical composition profiles of the SN ejecta from the explosion model, (2) the age of the SNR, (3) the amount 
of collisionless electron heating at the reverse shock, and (4) the density of the uniform AM. 
In Paper I, we analyzed the importance of (1) and (2); the impact of (3) and (4) is the focus of this Section.

\subsection{Collisionless Electron Heating at the Reverse Shock}

The unknown efficiency of collisionless electron heating in SNR shocks is one of the main uncertainties affecting 
the calculated X-ray spectra of SNRs. Direct application of the Rankine-Hugoniot relations at the shock front yields
\begin{equation}
  T_{p}=\frac{3m_{p}v_{s}^{2}}{16k}
\end{equation}
for each population of particles $p$, where $m_{p}$ is the particle mass, $v_{s}$ is the shock velocity and $k$ is Boltzmann's 
constant. Because of the large difference between electron and ion masses, the electrons are expected to be much colder than the 
ions, and the quotient of postshock specific internal energies defined as 
\begin{equation}
  \beta\equiv\frac{\varepsilon_{e,s}}{\varepsilon_{i,s}}=\frac{\overline{Z_{s}}T_{e,s}}{T_{i,s}}
\end{equation}
is expected to be close to 0, where $\overline{Z_{s}}$ is the preshock ionization state (i.e., the number of free electrons per ion
in the unshocked ejecta). However, \citet{cargill88:e_heating} argued that plasma waves 
can redistribute energy among cold electrons and hot ions at the shock, bringing the value of $\beta$ close to $\overline{Z_{s}}$
\citep[for a discussion of collisionless electron heating see][]{laming00:e_heatin_snr_shock}.  

So far, the observational evidence hints at a decreasing level of thermal equilibration with increasing shock speeds or Mach
numbers in the forward shocks of SNRs \citep[see][and references therein]{rakowski03:DEML71}. In the forward shock of
Tycho, \citet{ghavamian01:balmer_SNRs} found a value of $T_{e}/T_{i}\leq 0.1$ by analyzing the optical Balmer
emission, while \citet{vink03:sn1006-tequil} estimated a much lower value at the forward shock of SN\,1006 from X-ray
observations. The only constraint on the value of $\beta$ in the reverse shock of a young Type Ia SNR comes from the
absorbed spectrum of the Schweizer and Middleditch star behind SN\,1006, where the amount of thermal energy deposited in the
electrons was found to be negligible \citep{hamilton97:UV_SMstar_SN1006}. The model spectra presented in Paper I were calculated
assuming no collisionless electron heating at the reverse shock, effectively setting $\beta$ to the lowest possible value,
$\beta_{min}=\overline{Z_{s}}\cdot m_{e}/\overline{m_{i}}$, where $\overline{m}_{i}$ is the average ion mass in a fluid element.
It is clear from the works cited above that, while full thermal equilibration between ions and electrons at the shock
(i.e., $\beta=\overline{Z_{s}}$) is not compatible with the observations, values of $\beta$ larger than $\beta_{min}$ cannot be 
excluded.

The effect of varying amounts of collisionless electron heating at the forward shock on the X-ray spectrum emitted by the shocked AM
of SNRs in the Sedov stage was discussed in \citet{borkowski01:sedov}; here, we shall analyze the impact of a small
(but nonzero) amount of collisionless electron heating at the reverse shock on the properties of the shocked ejecta.
We will illustrate this effect using a delayed detonation model as an example. Of all the classes of one-dimensional 
Type Ia explosion models, delayed detonations have been the most successful in reproducing the light curves and spectra of Type Ia 
SNe \citep{hoeflich96:nosubCh}, and therefore it is of much interest to analyze the details of the X-ray emission that these 
models predict for Type Ia SNRs. Within the delayed detonations, we chose model DDTe because it has the largest amount of
intermediate mass elements (Si, S, etc.) in the ejecta. This should make it easier to estimate the effects that collisionless
electron heating at the reverse shock has on the prominent X-ray lines from these elements. 

\begin{figure} 

  \epsscale{0.80}
 
  \plotone{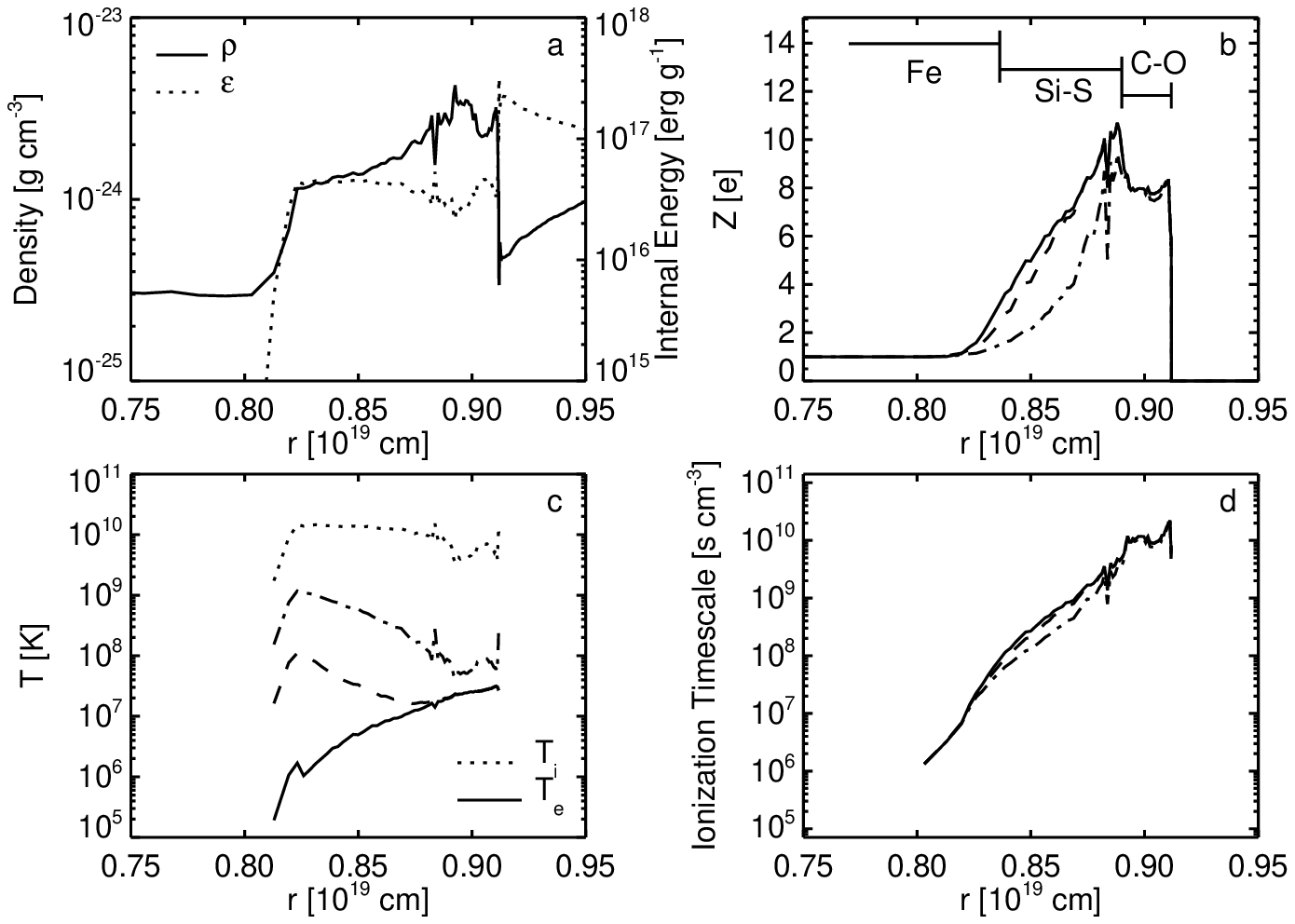}

  \caption{
    Shocked ejecta structure vs. radius for model DDTe interacting with an ambient medium of density
    $10^{-24}\,\mathrm{g\cdot cm^{-3}}$, 430 years after the explosion. The panels show the radial distribution of
    density and specific internal energy (a), mean number of electrons per ion, $\overline{Z}$, 
    with an indication of the ejecta layers dominated by Fe, Si-S and C-O (b), electron and ion temperatures (c) 
    and ionization timescale (d). The positions of the reverse shock and contact discontinuity are outlined by the limits of 
    the temperature plots in panel c. The three plots for $\overline{Z}$ (panel b), $T_{e}$ (panel c) and $\tau$ (panel d) represent 
    the values obtained with $\beta=\beta_{min}$ (solid), $\beta=0.01$ (dashed) and $\beta=0.1$ (dash-dotted).\label{fig-1}}

\end{figure}

Figure \ref{fig-1} shows the shocked ejecta  of model DDTe for a uniform 
AM of density $\rho_{AM}=10^{-24}\,\mathrm{g\cdot cm^{-3}}$, 430 yr after the explosion (the age of the Tycho SNR).
The electron heating and plasma ionization processes in the shocked ejecta have been calculated for $\beta=\beta_{min}$, $0.01$, 
and $0.1$. The unshocked ejecta were assumed to be singly ionized in all cases. As discussed in Paper I, the interaction of ejecta with 
the AM leads to the formation of density structures within the shocked ejecta, which strongly affect the distributions of 
ionization states (represented here by the average ion charge $\overline{Z}$), electron temperatures $T_{e}$, and ionization 
timescales ($\tau=\int n_{e}dt$). Together with the chemical composition profile of the ejecta, these distribution 
functions determine the spectral properties of each element, and ultimately the shape of the emitted X-ray spectrum. 
An important feature of the shocked ejecta is the pronounced density peak towards the contact
discontinuity (CD) that appears in all Type Ia SNR models \citep[see Paper I and][]{dwarkadas98:typeIa}.
 
In the case with no collisionless heating ($\beta=\beta_{min}$), the electron temperature profile rises monotonically from the 
reverse shock to the CD, as internal energy is gradually redistributed from the hot ions to the cold electrons through Coulomb 
collisions. The electron temperature profile peaks at the CD, where the fluid elements have been shocked for the longest time and 
have the highest density (the rate at which the ion and electron temperatures equilibrate in the shocked ejecta scales 
with $\rho$, see eq. 1 in Paper I). Increasing the value of $\beta$ makes the electrons just behind the reverse 
shock hotter, but the electron temperature drops as numerous cold electrons are liberated in the ongoing ionization process and the 
total internal energy in the electrons is redistributed among more particles. For $\beta=0.01$, the electron temperature profile 
eventually relaxes to the profile obtained without any collisionless heating at the reverse shock, but in the $\beta=0.1$ case 
there is a significant residual temperature excess even in the outermost ejecta layers. The average ionization state and ionization 
timescale become severely affected only for $\beta \ge 0.1$, when the electrons reach extremely high($\sim 10^{9}$ K) temperatures 
behind the reverse shock. The ionization process is less efficient at these extreme temperatures, leading to lower mean ion 
charges and ionization timescales in the shocked ejecta. 

\begin{figure}

  \epsscale{0.80}

  \plotone{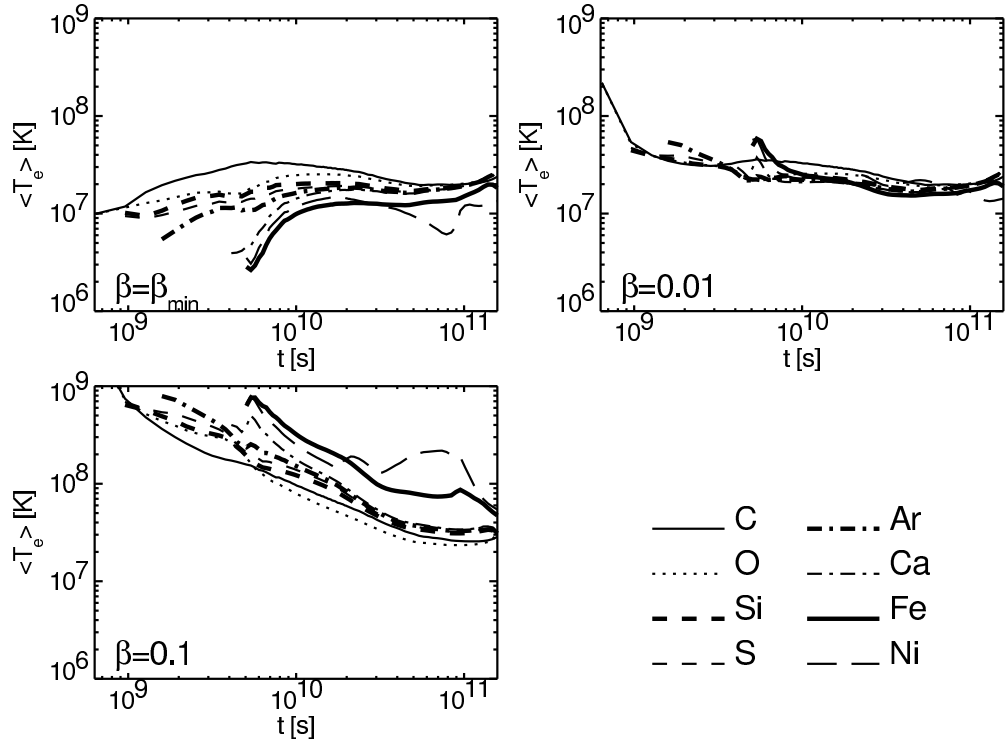}

  \caption{
    Evolution of $\langle T_{e} \rangle _{X}$ for C, O, Si, S, Ar, Ca, Fe and Ni in the shocked ejecta of model DDTe, with 
    $\rho_{AM}=10^{-24}\,\mathrm{g\cdot cm^{-3}}$. The top left panel corresponds to $\beta=\beta_{min}$, and is the
    same as panel b of Figure 5 in Paper I, but with a different scale. The top right panel corresponds to $\beta=0.01$, 
    and the bottom left panel to $\beta=0.1$.\label{fig-2}}

\end{figure}

The significance of this modification in the electron temperature profile is better understood when viewed in the context 
of the stratification that is inherent to 1D Type Ia SN explosion models. In the case of DDTe, as in all delayed detonation 
models, the inner ejecta layers are dominated by Fe and Ni, surrounded by a region rich in intermediate mass elements 
(mostly Si and S, but also Ar, Ca, and others), with O dominating the outermost ejecta layers. In Figure 
\ref{fig-1}, this stratified structure has been represented schematically in panel b. In this example,
the increase in $T_{e}$ caused by collisionless electron heating at the reverse shock affects primarily the Fe-rich ejecta
layers at this age. In terms of the emission measure-averaged electron temperature for each element X, 
$\langle T_{e}\rangle _{X}$, increasing the value of $\beta$ effectively reverses the approximate ordering in $T_{e}$ of the 
ejecta elements that is maintained throughout the evolution of the SNR. This is illustrated in Figure \ref{fig-2}, 
which can be contrasted with Figure 5 in Paper I. 

In all the calculations presented here and in Paper I, $\overline{Z_{s}}$ has been set to 1, which is generally a
good approximation for NEI plasmas in SNRs. Photoionization by UV starlight or by X-rays emitted by the shocked material in 
the SNR could raise $\overline{Z_{s}}$, but only by a factor of 3--4 \citep[see][]{hamilton88:reionization_fe_sn1006}. After 
the shock, the heavy element plasma ionizes rapidly, and the values of $\overline{Z}(t)$ in fluid elements that started with different
$\overline{Z_{s}}$ will converge over time. Due to this, the postshock X-ray emission is generally insensitive to moderate 
variations in the preshock ionization. In the presence of collisionless electron heating, increasing $\overline{Z_{s}}$ for a fixed 
value of $\beta$ raises the number of 'hot' electrons and decreases their temperature in the same proportion. X-ray spectra, however, 
are sensitive mostly to the total internal energy transferred to the electrons at the shock and to the final electron temperature, and 
not so much to how the internal energy is 
distributed among the electrons. For all the values of $\overline{Z_{s}}$, the final electron temperature is similar because the 
final value of $\overline{Z}$ is very similar. In view of this, we do not expect significant deviations from the results presented 
here for $\overline{Z_{s}}>1$ at a fixed value of $\beta$.

\subsection{The Ambient Medium Density}

The density of the AM affects the
spectral properties of the elements in the shocked ejecta in a dramatic way. This is due to two closely related effects: the 
acceleration of 
all the collisional plasma processes in denser media on one hand and the scaling of the hydrodynamic models with $\rho_{AM}$ 
on the other hand (see section 4.1 in Paper I for a discussion of the hydrodynamic scaling).

\begin{figure}

  \epsscale{0.80}

  \plotone{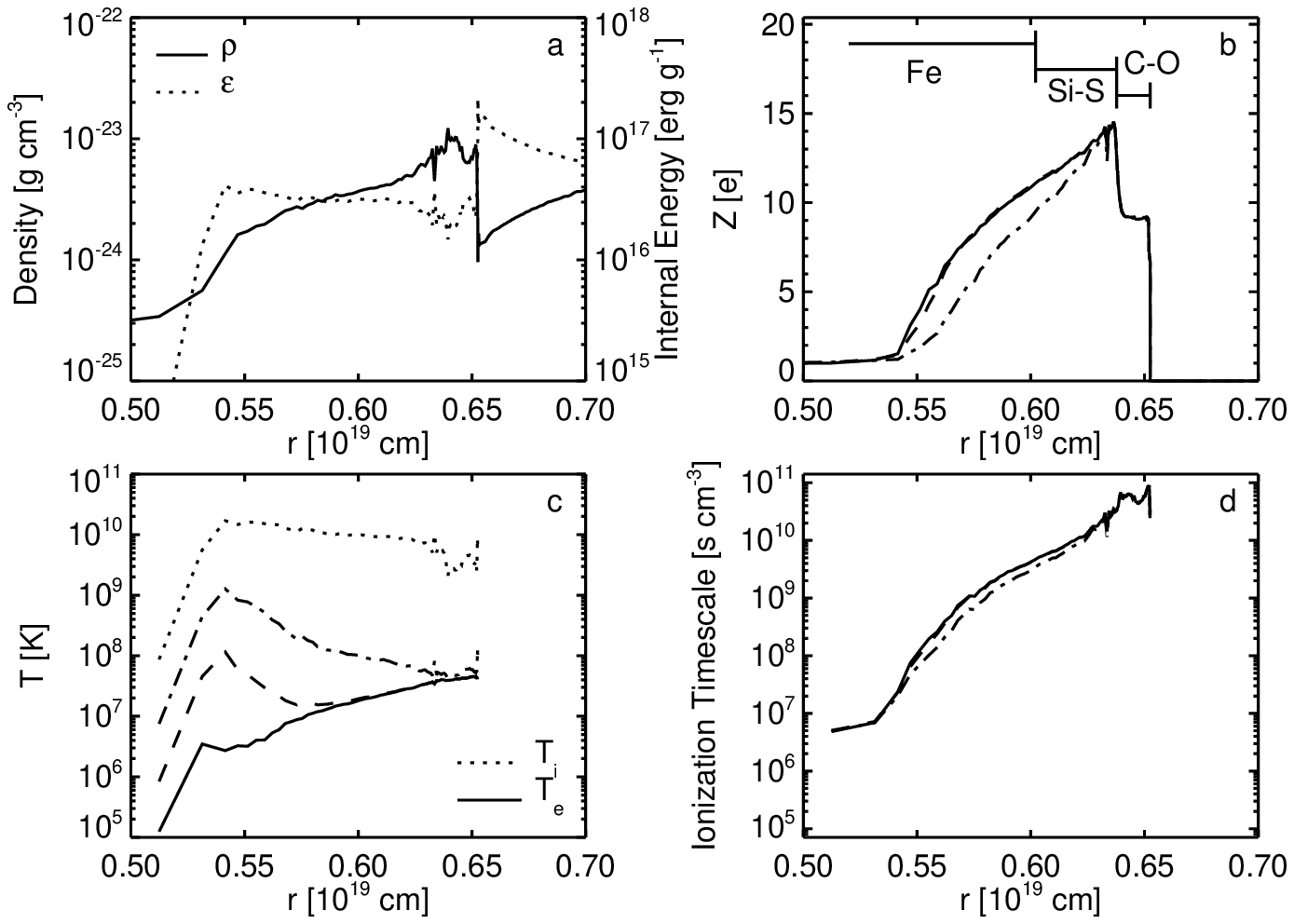}

  \caption{  
    Shocked ejecta structure vs. radius for model DDTe, with 
    $\rho_{AM}=5\cdot10^{-24}\,\mathrm{g\cdot cm^{-3}}$, 430 years after the explosion. See Figure \ref{fig-1} for an
    explanation of the plots and labels.\label{fig-3}}

\end{figure}

\begin{figure}

  \epsscale{0.80}

  \plotone{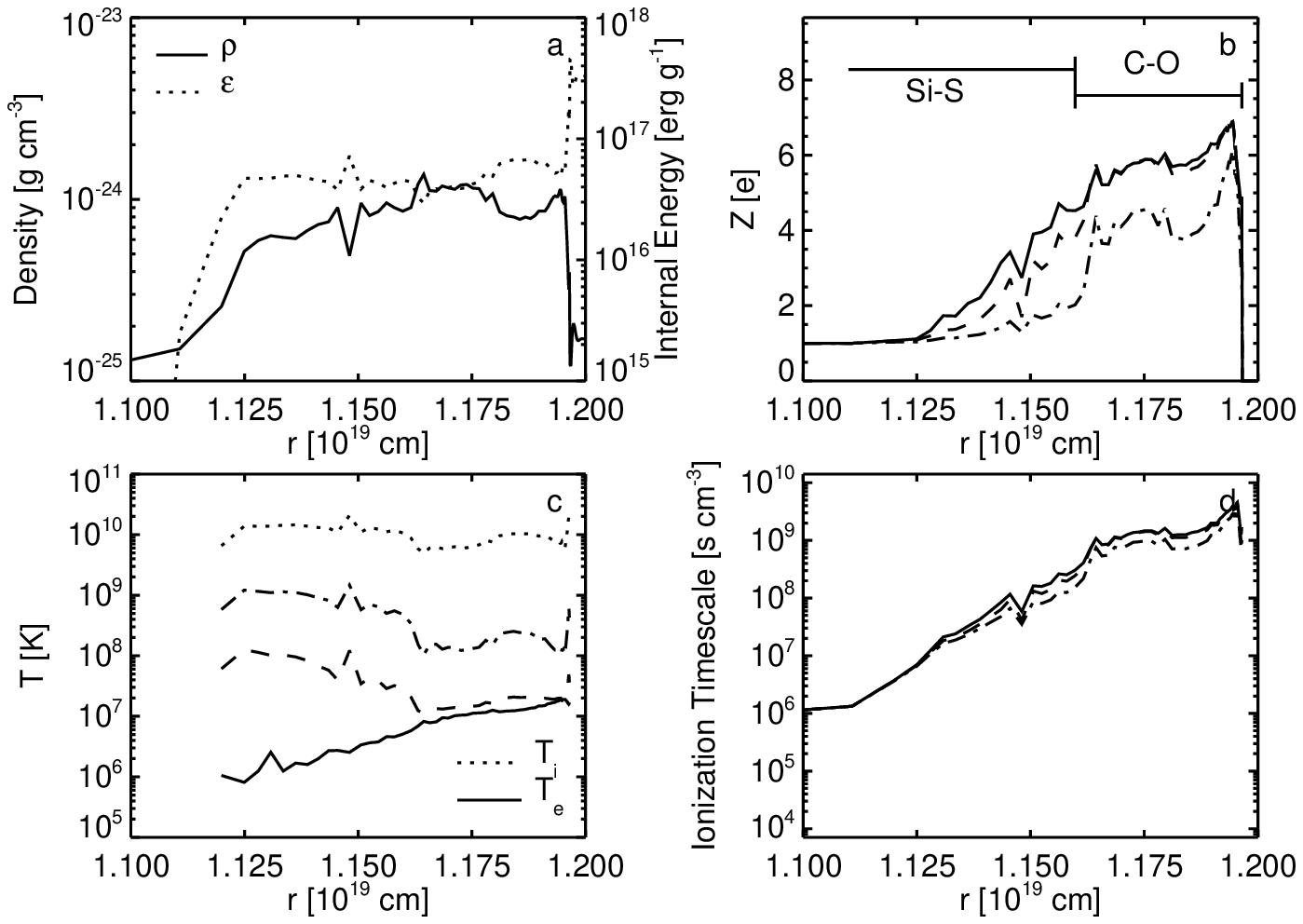}

  \caption{
    Shocked ejecta structure vs. radius for model DDTe, with 
    $\rho_{AM}=2\cdot10^{-25}\,\mathrm{g\cdot cm^{-3}}$, 430 years after the explosion. See Figure \ref{fig-1} for an
    explanation of the plots and labels.\label{fig-4}}

\end{figure}

These effects are illustrated in Figures \ref{fig-3} and \ref{fig-4}, which display the structure of the shocked 
ejecta of model DDTe at an age of 430 yr after the explosion, for a uniform AM of 
$\rho_{AM}=5\cdot10^{-24} \mathrm{g\cdot cm^{-3}}$ and $\rho_{AM}=2\cdot10^{-25} \mathrm{g\cdot cm^{-3}}$,
respectively. These simulations are also shown for $\beta=\beta_{min}$, $0.01$, and $0.1$ to facilitate 
comparison with Figure \ref{fig-1}. As a result of the scaling laws mentioned in Paper I, a SNR in a denser
AM will be in a more evolved evolutionary stage at any given time, and vice versa -- 
note how the reverse shock has not reached the Fe-dominated region of the ejecta at $t=430$ yr for 
$\rho_{AM}=2\cdot10^{-25} \mathrm{g\cdot cm^{-3}}$. The mean ionization state in the shocked ejecta, which peaks at 
$\overline{Z}\simeq10$ in the outermost Si-dominated layers for $\rho_{AM}=10^{-24} \mathrm{g\cdot cm^{-3}}$, 
rises as high as $\overline{Z}\simeq15$ for $\rho_{AM}=5\cdot10^{-24} \mathrm{g\cdot cm^{-3}}$ in the same region,
but only reaches $\overline{Z}\simeq6$ for $\rho_{AM}=2\cdot10^{-25} \mathrm{g\cdot cm^{-3}}$. These differences in 
the mean ionization state correspond to differences in the ionization timescales of roughly an order of magnitude throughout
the shocked ejecta for each factor 5 increase in $\rho_{AM}$. The electron temperature profiles are affected as well, 
although only by a factor of $\backsim2-3$. The higher densities and faster ionization rates tend to mitigate the effect of 
collisionless electron heating, favoring the convergence to the canonical $\beta=\beta_{min}$ case (compare panels c 
of Figures \ref{fig-3} and \ref{fig-4}).

Varying the value of $\rho_{AM}$  has an immediate impact on the emission measures and emission measure averaged quantities, mainly 
through the hydrodynamic scaling laws mentioned in Paper I. The approximate scaling of $EM_{X}(t)$  and  $\langle \tau \rangle _{X}(t)$  
is given by eqs. (3)-(5) in Paper I:  $\rho_{AM}^{-1/3}$  for the  t  axis,  $\rho_{AM}$  for  $EM_{X}(t)$  and  
$\rho_{AM}^{2/3}$  for  $\langle \tau \rangle _{X}(t)$. These approximate scalings are accurate within a factor of 2 for  
$2\cdot10^{-25}\leq\rho_{AM}\leq5\cdot10^{-24} \mathrm{g\cdot cm^{-3}}$, but they might break down for values of  
$\rho_{AM}$  outside this range. The effect of a change of $\rho_{AM}$  on the electron temperatures is more complex, and 
difficult to approximate with sufficient accuracy in view of the sensitivity of X-ray spectra to electron temperature.
Whereas an accuracy within a factor of 2 is reasonable for $EM_{X}$  and  $\tau_{X}$ , which span several orders of magnitude, 
changes by a factor of 2 are too large compared with the more modest (but still up to 2 orders of magnitude) range in $T_{e}$.

\subsection{Effects on the X-ray Spectrum}

\begin{figure}

  \epsscale{0.60}

  \plotone{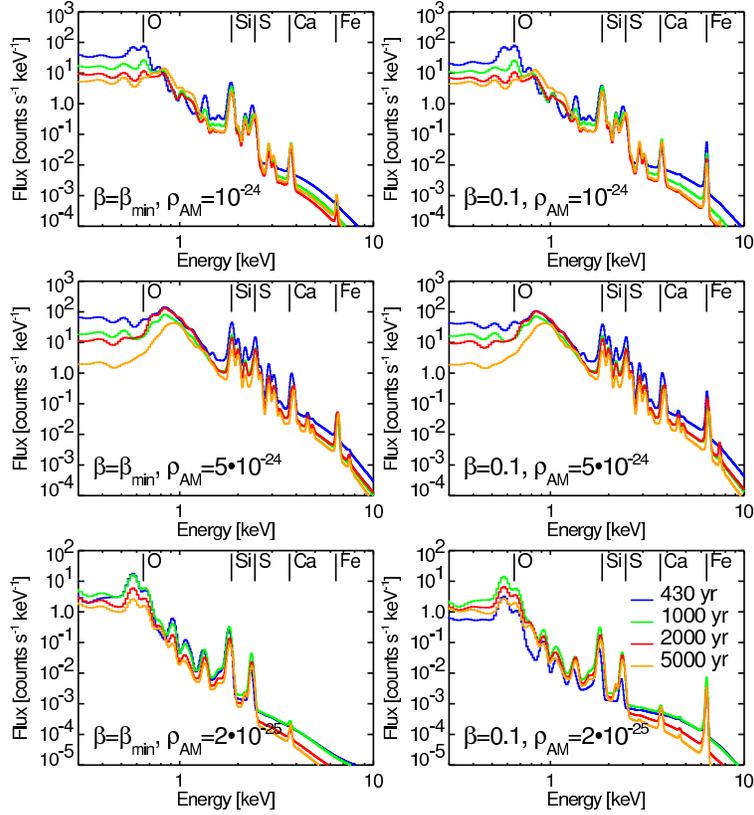}

  \caption{
    Unabsorbed X-ray spectra from the shocked ejecta of the DDTe model, 430, 1000, 2000 and 5000 yr after the SN explosion,
    for $\rho_{AM}=10^{-24} \mathrm{g\cdot cm^{-3}}$ (top panels), $5\cdot10^{-24} \mathrm{g\cdot cm^{-3}}$ 
    (middle panels), and $2\cdot10^{-25} \mathrm{g\cdot cm^{-3}}$ (bottom panels), convolved with the spectral response of the 
    {\it XMM-Newton} EPIC MOS1 CCD camera. The spectra in the left panels correspond to $\beta=\beta_{min}$, those in 
    the right panels to $\beta=0.1$. The K$\alpha$ lines of Fe, Ca, S, and Si, as well as the O Ly$\alpha$ line, have been 
    marked for clarity, and fluxes are calculated at a fiducial distance of 10 kpc. Note that the spectral code has no atomic data 
    for Ar. See the on-line edition for a color version of this Figure.\label{fig-5}}

\end{figure}

In Figure \ref{fig-5}, the temporal evolution of the spectra from the shocked ejecta of model DDTe is presented for three 
values of  $\rho_{AM}$ ( $10^{-24}$, $5\cdot10^{-24}$,
and  $2\cdot10^{-25} \mathrm{g\cdot cm}^{-3}$) and two values of  $\beta$ ($\beta_{min}$  and $0.1$). A preliminary 
inspection reveals that variations in $\rho_{AM}$  have profound effects on the calculated spectra. As expected from the 
scaling law for $<\tau>_{X}$, the plasma ionization state varies greatly, and the presence of different ions   
results in emission of different ionic lines. At  $\rho_{AM}=5\cdot10^{-24} \mathrm{g\cdot cm^{-3}}$, 
for instance, the more advanced ionization state of Fe leads to a higher flux in the Fe L complex, which blends with
O Ly$\alpha$ and Mg He$\alpha$  emission at the {\it XMM-Newton} CCD spectral resolution. The increase in the Fe K$\alpha$  line, 
on the other hand, is due to the higher temperatures in the ejecta. The prominent O He$\alpha$  line at 0.56 keV, seen at 
early times for $\rho_{AM}=10^{-24} \mathrm{g\cdot cm^{-3}}$,  disappears at higher values of  $\rho_{AM}$, because He-like 
O is ionized more rapidly. The overall higher ionization state of the plasma also leads to an increase in the flux of the 
Ly$\alpha$ lines of Si and S, and a shift towards higher energies of the Ca K$\alpha$ line. The shape and flux of the continuum 
emission also change. At lower densities, these effects are reversed. The Fe K$\alpha$ line and Fe L complex virtually 
disappear, revealing the underlying Ne He$\alpha$  and Ne Ly$\alpha$  lines at 0.9 and 1.0 keV. The O He$\alpha$ line becomes more 
important than O Ly$\alpha$, and the Ly$\alpha$ and He$\beta$ lines of Si and S vanish almost completely, as well as the 
Ca K$\alpha$ line. The continuum flattens and the emitted flux is generally lower at all energies.

In contrast with the global effects of variations of $\rho_{AM}$, changes in the amount of collisionless heating at the 
reverse shock have a different impact on different elements in a model with stratified ejecta like DDTe. For 
$\rho_{AM}=10^{-24} \mathrm{g\cdot cm^{-3}}$, the flux in the Fe K$\alpha$  line, which probes material at higher  
$T_{e}$ and lower $\tau$ than the Fe L complex, is increased by almost two orders of magnitude for $\beta = 0.1$. 
None of the other elements seems to be affected at this AM density, although model DDTe has a significant amount of S, Si 
and Ca in the inner ejecta. This increase in the Fe K$\alpha$ flux becomes less pronounced with time, and is accompanied 
by a slight change in the shape of the continuum. For $\rho_{AM}=5\cdot10^{-24}\, \mathrm{g\cdot cm^{-3}}$, the continuum is 
unaffected and the increase of the Fe K$\alpha$ line flux is reduced to less than an order of magnitude at early times, 
disappearing completely at late times. At $\rho_{AM}=2\cdot10^{-25} \mathrm{g\cdot cm^{-3}}$, however, the collisionless 
electron heating has a more noticeable effect. The shape of the spectrum is not changed at low energies, but the flux is 
somewhat lower at early times for $\beta=0.1$. At high energies, the level of continuum rises and the flux in the 
Fe K$\alpha$ line flux greatly increases. The effects of collisional electron heating can be clearly seen even at CCD 
spectral resolution. With a higher spectral resolution such as provided by {\it ASTRO-E2}, the predicted large temperature 
increases caused by collisionless heating (Fig. \ref{fig-2}) should be detectable through various temperature-sensitive 
line diagnostics for a number of different chemical elements within the shocked ejecta.

We emphasize that model DDTe is presented here just as an illustrative example. For obvious reasons, it is not 
practical to present the effects of $\rho_{AM}$ and $\beta$ on the spectra of all the models in our grid. 
Although the details may vary, the general trends identified here for DDTe can be applied to most of the other models
\citep[for a discussion of other delayed detonation models, see][]{badenes04:model_grid}.

To conclude, we note that collisionless electron heating at the reverse shock can have interesting effects on the spatially
resolved X-ray emission. In particular, the enhanced flux in the Fe K$\alpha$ line discussed above
would come mainly from the hotter regions of Fe-rich ejecta close to the reverse shock (see the shape of the electron temperature 
profile in Figures \ref{fig-1}, \ref{fig-3}, and \ref{fig-4} for values of $\beta$ above 0.01). This scenario is compatible, at 
least qualitatively, with the finding that the Fe K$\alpha$ emission peaks at a smaller radius than the Fe L and Si He$\alpha$ 
emission in the X-ray CCD images of both the Tycho \citep{hwang98:tycho-Feemission} and Kepler \citep{cassam03:kepler} SNRs.
Collisionless electron heating provides a simpler explanation to the rise of the electron temperature profile
towards the reverse shock than the relic of an interaction with a circumstellar medium invoked by \citet{dwarkadas98:typeIa} 
for the Tycho SNR.

\section{X-RAY SPECTRUM FROM THREE DIMENSIONAL TYPE Ia EXPLOSION MODELS} \label{sec:Expanding}

\subsection{Type Ia SN Explosions in 3D: Fundamental Properties} \label{subsec:3Dmodels}

In Paper I, we introduced a grid of eight one dimensional Type Ia SN explosion models that included examples of all the
paradigms currently under debate: sub-Chandrasekhar explosions, deflagrations, delayed detonations, and pulsating delayed 
detonations. This reduced grid is just a representative sample of a more extensive grid of 19 models, which constituted
the base for the study of Type Ia SNRs conducted by \citet{badenes04:PhD}. The remaining 11 grid models are intermediate 
cases obtained by varying the parameters involved in the calculation of each explosion paradigm. We have included these models in 
the Appendix, both for reference in future works and for the convenience of those readers who want to use our synthetic SNR spectra 
for their own research. This grid is one representation of our current understanding of one dimensional Type Ia explosion models, 
upon which most of our knowledge of the physics of Type Ia SNe is based. 

In view of the recent developments in the field, however, it has become clear that 1D calculations will soon be superseded by 
the three dimensional models that have begun to appear in the literature
\citep{reinecke02:Ia3D,gamezo03:Ia3D,travaglio04:3D,garcia-senz03:Ia3D}. These works have focused on pure deflagrations in 3D, 
proving that they are capable of producing robust explosions, but the ability of these models to explain the observations of 
Type Ia SNe has not been fully established yet. A common feature in all 3D deflagration models, and the most remarkable difference 
with respect to 1D models, is the uniform mixing of unburnt C and O material with $^{56}$Ni and the other products of nuclear 
burning throughout the ejecta. This mixing is due to the deformation of the flame front caused by Rayleigh-Taylor instabilities, 
an effect which seems unavoidable in 3D deflagrations. There has been some concern that the presence of large amounts of C and O 
in the inner 
layers of ejecta would lead to a spectral evolution inconsistent with optical observations \citep{gamezo03:Ia3D}, but complex 
spectral simulations are required to verify this claim \citep{baron03:detectability}. Alternatives to the 3D deflagration scenario 
are being explored right now, including delayed detonations in 3D \citep{garcia-senz03:DDT3D,gamezo04:DDT3D,gamezo05:DDT3D} and 
two new explosion paradigms: gravitationally confined detonation \citep{plewa04:gcd} and pulsating reverse detonation 
\citep{bravo05:SNIa_3D_review}, 
but none of these models has been completely understood yet. Here, we study 3D deflagration models in the context of the X-ray spectra 
of young SNRs to provide an independent method of assessing their viability for Type Ia explosions.  

\begin{figure}

  \centering
  \includegraphics[scale=0.8]{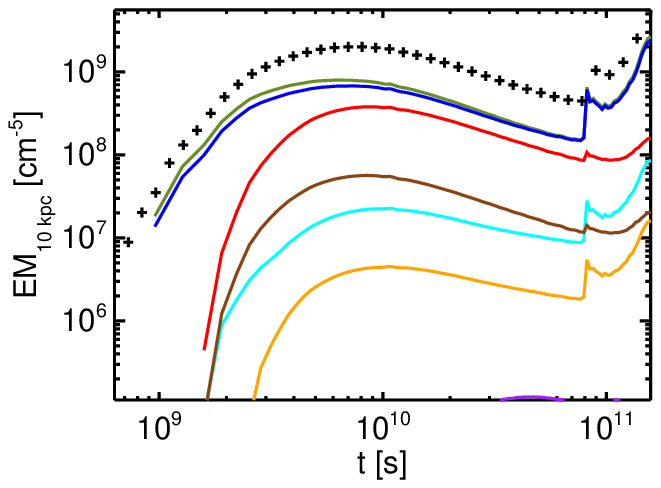}
  \includegraphics[scale=0.8]{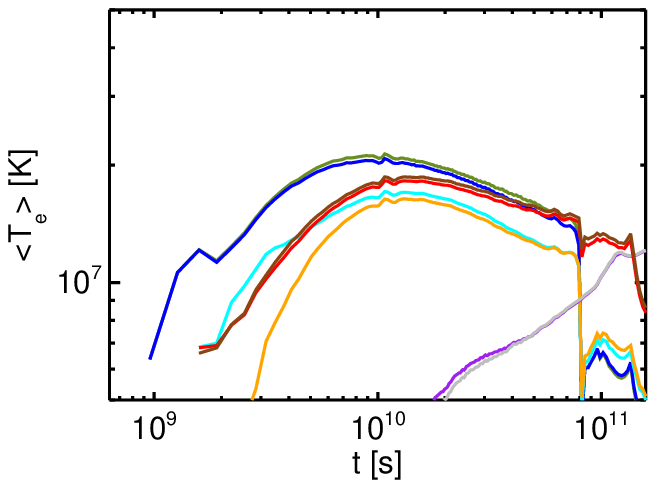}

  \centering
  \includegraphics[scale=0.8]{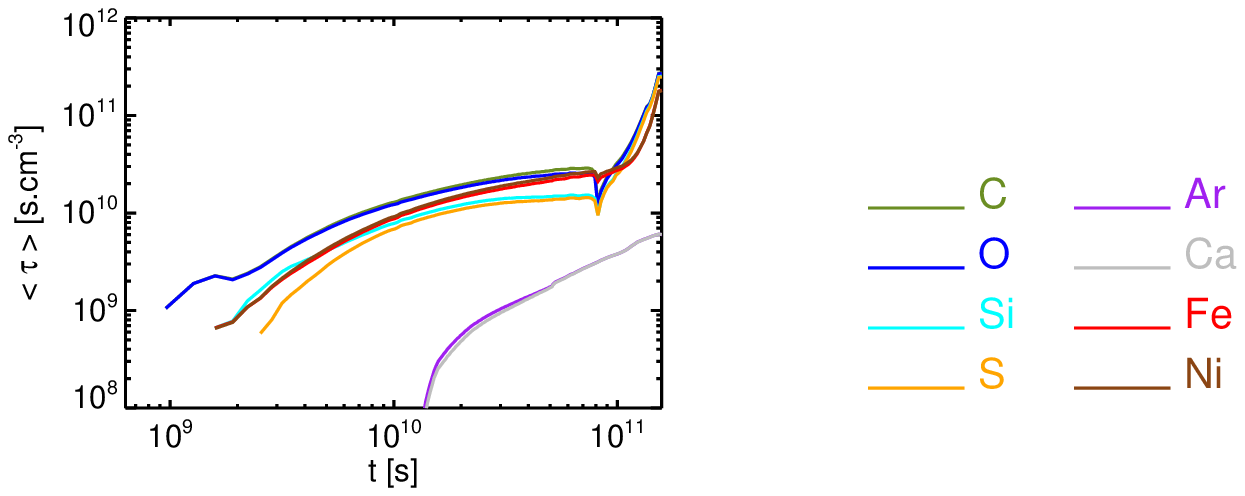}

  \caption{
    $EM(t)$ (top left), $<T_{e}>(t)$ (top right) and $<\tau>(t)$ (bottom) for the principal 
    elements in the shocked ejecta of model B30U, interacting with a uniform AM of $\rho_{AM}=10^{-24} \mathrm{g\cdot cm^{-3}}$. 
    The crosses in the $EM(t)$ plot represent the total emission measure of the shocked ejecta. See the on-line
    edition for a color version of this Figure. \label{fig-6}}

\end{figure}

We use a one dimensional average of model B30U, a 3D deflagration from \citet{garcia-senz03:Ia3D}, to illustrate what can be expected 
from this class 
of models. The chemical composition and density profile of this model are presented in the Appendix, and they are 
very similar to those of the models obtained by \citet{gamezo03:Ia3D} and \citet{travaglio04:3D}, even though
the computational techniques and the resolution of the calculations are different in each case 
\citep[see Table 1 in][for a more detailed comparison of these works]{bravo05:SNIa_3D_review}. This shows that three dimensional 
deflagrations are relatively well understood, and supports our use of model B30U as a
representative example of this class. The evolution 
of the emission measures and emission measure-averaged ionization timescales and electron temperatures of the principal elements
in the ejecta of B30U are shown in Fig. \ref{fig-6} for an interaction with $\rho_{AM}=10^{-24} \mathrm{g\cdot cm^{-3}}$ 
and $\beta=\beta_{min}$. Interpretation of differences between these plots and Figures 4, 5, and 6 of Paper I is not 
straightforward, because the 3D calculations
are not fully self-consistent with the 1D models of the grid (the effect of the energy deposited by the decay of $^{56}$Ni 
on the density profile, for instance, has not been taken into account). Nevertheless, the main features of the evolution of
the shocked ejecta do not depend on such details. The most striking property of this model is the similarity in the spectral 
properties ($\langle T_{e} \rangle$ and $\langle \tau \rangle$) of Fe and Si throughout the evolution of the SNR. This is in marked 
contrast to 1D models, where the stratification of the ejecta leads to significant differences between Fe and Si. The abrupt 
changes in several plots seen at $t\sim 8\cdot10^{10}$ s are due to the impact of the reverse shock on a remnant of unburnt 
white dwarf material formed in model B30U \citep[for details, see][]{garcia-senz03:Ia3D}.

\begin{figure}

  \epsscale{0.60}

  \plotone{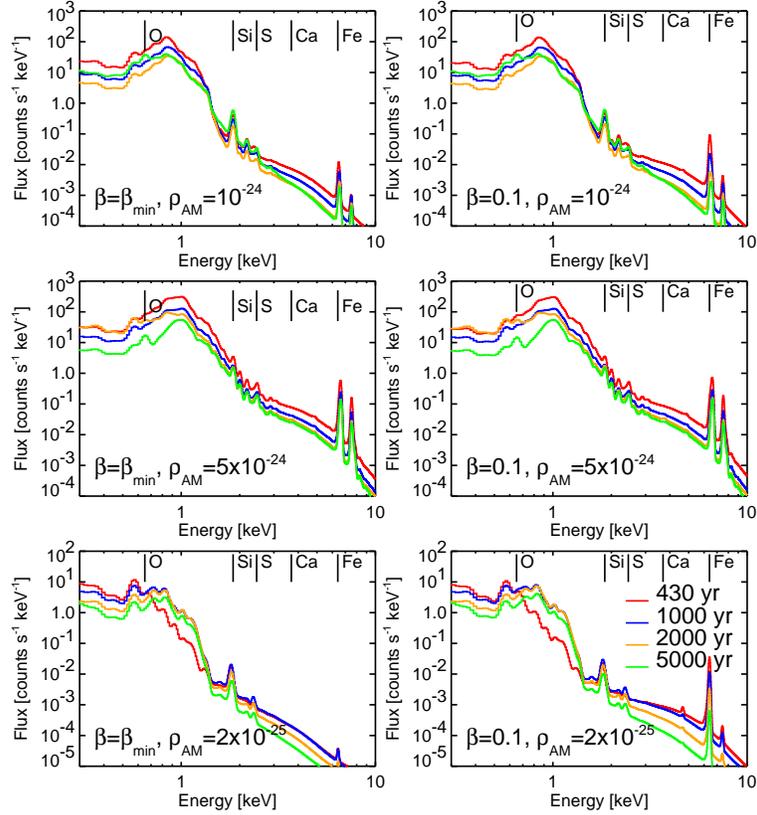}

  \caption{
    Unabsorbed X-ray spectra from the shocked ejecta in the B30U model, 430, 1000, 2000 and 5000 yr after the SN explosion,
    for $\rho_{AM}=10^{-24} \mathrm{g\cdot cm^{-3}}$ (top panels), $5\cdot10^{-24} \mathrm{g\cdot cm^{-3}}$ 
    (middle panels), and $2\cdot10^{-25} \mathrm{g\cdot cm^{-3}}$ (bottom panels). The spectra in the left panels correspond to
    $\beta=\beta_{min}$, those in the right panels to $\beta=0.1$. See Figure \ref{fig-5} for an explanation of the labels and
    plots. The online edition of the journal contains a color version of this Figure. \label{fig-7}}

\end{figure}

In Figure \ref{fig-7}, we show the ejecta spectra of model B30U at the same values of $t$, $\rho_{AM}$ and $\beta$ as in Figure 
\ref{fig-5} for model DDTe. The most remarkable properties of these spectra are the high Fe L-shell flux and the presence of the 
prominent Ni K$\alpha$ line at $\sim 7.5$ keV (except at the lowest AM densities). This is due to the large amounts of Fe and Ni 
that are found in the outermost layers of B30U, where the density of the shocked ejecta is highest. The results are a long Fe
ionization timescale, which leads to the enhanced Fe L-shell flux, and a high Ni emission measure leading to a strong 
Ni K$\alpha$ emission. Such a strong Ni K$\alpha$ line has never been observed in thermal X-ray spectra of SNRs. Another interesting
feature is the relative weakness of the Si and S K$\alpha$ lines. The reason for this is twofold: first, 3D deflagrations  
produce smaller amounts of Si, S and other intermediate mass elements than the conventional 1D delayed detonations; second, 
equivalent widths of the Si and S lines are smaller because of the strong continuum produced by the large amount of C and O
that is present throughout the ejecta. These spectral characteristics exhibited by model B30U are common to all 3D deflagrations
with well mixed ejecta.

\subsection{Comparison with X-ray observations of SNRs} \label{subsec:Can3Ddoit?}

We compare the results of our simulations for the 3D deflagration model B30U with the basic properties of Type Ia 
SNRs. The prediction 
of similar emission measure-averaged electron temperatures and ionization timescales for Si and Fe can be easily tested by examining 
X-ray observations. We have searched the literature for young SNRs with published good-quality X-ray spectra that have been 
classified as Type Ia. Six objects meet these requirements: the historical remnants of Tycho, Kepler and SN\,1006, and three 
Large Magellanic Cloud SNRs: N103B, 0509-67.5, and DEM L71. We note that the classification of Kepler's SNR as Type Ia is 
controversial \citep[e.g.,][]{blair04:kepler}. The remnant of SN\,1006 is not suitable for our 
purposes because it lacks strong Fe emission \citep{koyama95:acc_elect_SN1006}. In the case of DEM L71, although the X-ray spectrum 
of this SNR has been analyzed in some detail \citep[see][]{hughes03:DEML71,vanderHeyden03:DEML71}, we found no published estimates 
of temperatures and ionization timescales for Fe and Si in the ejecta. The results of the analysis of the integrated spectrum 
for the other four SNRs are summarized in Table \ref{tab-1}. 

The spectral properties of all the SNRs considered here show that an important fraction of the Fe in the shocked ejecta is emitting 
under conditions different from those of the Si. The authors of the works referenced in Table \ref{tab-1} accounted for this by either 
adding a spectral component made of pure Fe to their fits or by using plane-parallel models that allowed to treat Fe and Si 
separately by assigning different values of $n_{e}t$ and $kT$ to each element. Since the analysis techniques, models, and data 
quality were different in each case, these results can only be compared either with our models or among themselves in a qualitative 
way. Nevertheless, a clear trend can be observed in all four SNRs considered here: the Fe component was always 
hotter than the Si component by at least a factor of 2. The Fe was at a lower ionization timescale in three out of four objects: 
Tycho, Kepler, and N103B. In 0509-67.5, however, the Fe component has a higher ionization timescale than Si. In this case, 
the statistics of the Fe K$\alpha$ line were poor, and the ionization timescale of Fe was constrained mostly by fitting
the Fe L complex. \citet{warren03:0509-67.5} note that their fit to the Fe L complex emission was not complete, because a strong 
line had to be added by hand. Improved atomic physics and higher resolution data would be highly desirable to confirm this 
result for 0509-67.5. 

Since the emission measure averaged ionization timescales and electron temperatures of Si and Fe do not differ by more than 30\% 
in model B30U (see Figure \ref{fig-6}), we conclude that this model is in conflict with the observations listed on Table 
\ref{tab-1}, at least within the limitations of our simulations. This conclusion is extensible to any model in which Fe and Si 
are well mixed throughout the ejecta, and therefore can be applied to all the 3D deflagration models for Type Ia SNe discussed in 
the previous section. As we have seen in \S~\ref{sec:Techniques}, a plasma state with
$\langle T_{e} \rangle_{Si}\,<\, \langle T_{e} \rangle_{Fe}$ and 
$\langle \tau \rangle_{Si}\,>\,\langle \tau \rangle_{Fe}$ arises naturally in Type Ia SN models with stratified ejecta, 
such as 1D delayed detonations or pulsating delayed detonations, that undergo a moderate amount of collisionless electron heating 
at the reverse shock. The ionization timescales of Fe and Si in 0509-67.5 are clearly incompatible with this 
scenario, but in this case \citet{warren03:0509-67.5} found a very low amount of Fe in the shocked ejecta, with Fe to Si 
abundance ratios below 0.07. While this is very difficult to interpret in the context of well-mixed Type Ia SN ejecta, it could be more
easily explained if the reverse shock were just entering the Fe-dominated region in stratified ejecta. A detailed
comparison of our models with this SNR would be required to confirm this hypothesis.

We emphasize that our simulations based on one dimensional averages are too simple to rule out well-mixed 3D Type Ia SN explosion 
models. We do not account for a number of processes that might result in the Fe and Si in the ejecta emitting under different 
conditions, like the Ni bubble effect \citep{basko94:Nibubbles-SN87A,blondin01:dynam_fe_bubbles_SNRs} or the formation of clumps
in the ejecta \citep{wang01:Ia-inst-clump}. Nevertheless, we find that the observations of Type Ia SNRs seem easier to explain in the 
light of Type Ia SN explosion models with stratified ejecta.

\section{DISCUSSION AND CONCLUSIONS} \label{sec:Discussion}

In this paper, we have examined several important aspects of the X-ray spectral models for the ejecta in Type Ia SNRs 
that were introduced in \citet{badenes03:xray}. We have explored the impact of the amount of collisionless electron heating 
at the reverse shock, $\beta$, and the density of the ambient medium, $\rho_{AM}$, on the integrated X-ray emission from the 
ejecta in Type Ia SNR models of different ages. We found that even small amounts of collisionless electron heating can modify 
the electron temperature profile inside the ejecta in a significant way, leading to a region of hot material
at low ionization timescales close to the reverse shock. In the context of Type Ia SN explosion models with stratified 
ejecta, this modified temperature profile can affect the emission from the inner layers rich in Fe for a broad range of dynamical 
ages, increasing the flux in the Fe K$\alpha$ complex. This could explain why the Fe K$\alpha$ emission
peaks at smaller radii than Fe L in both the Tycho \citep{hwang98:tycho-Feemission} and Kepler \citep{cassam03:kepler} SNRs.
The density of the AM also has a strong impact on the X-ray emission from
the ejecta. For higher values of $\rho_{AM}$, the SNR is in a more advanced evolutionary stage at a given age, and the 
ionization timescale of the shocked ejecta increases significantly. At lower values of $\rho_{AM}$, the ionization 
timescales decrease and the evolutionary stage is less advanced. We have provided approximate scaling laws to estimate 
these effects, and discussed their impact on specific emission lines and line complexes through an example.

We have also reviewed the fundamental properties of the recent deflagration models for Type Ia SNe calculated in 3D, and
their performance in the context of the X-ray spectra of SNRs. Using our 1D simulation scheme, we have shown that the 
mixing of fuel and ashes throughout the ejecta, which is a common feature of these 3D explosion models, results in all the
elements in the shocked ejecta of the SNR having very similar spectral characteristics. In particular, the emission measure 
averaged ionization timescales and electron temperatures of elements like Fe and Si are always very close to each other. This 
is in conflict with the observations of Type Ia SNRs in our Galaxy and the Magellanic Clouds, where the Fe and Si in the 
shocked ejecta are found to be emitting under different physical conditions. Within the limitations of our 1D simulation
scheme, these observations are easier to explain with Type Ia explosion models that have stratified ejecta than with models
that have well mixed ejecta like 3D deflagrations.  

We believe that our models represent a significant improvement over current methods of analyzing and interpreting the
X-ray emission from the shocked ejecta in SNRs. In order to facilitate the comparison between our models and observations, 
we have built a library of synthetic spectra, which is available from the authors upon request. This library is presented 
in the Appendix, where more Type Ia SN explosion models are introduced, and some aspects relevant to the comparison
between the synthetic spectra and observations are discussed. A detailed example of this kind of comparison in the framework 
of the ejecta emission from the Tycho SNR will be the subject of a forthcoming paper (Badenes et al., in preparation).

\acknowledgements
We wish to thank Jack Hughes and Jessica Warren for detailed discussions concerning 0509-67.5. We also acknowledge conversations 
with Una Hwang and Martin Laming on several aspects of the research presented here. We are grateful to the anonymous referee for 
suggestions that helped to improve the quality of this manuscript. This research has been partially supported by the CIRIT and 
MCyT in Spain, through grants AYA2000-1785, AYA2001-2360, and AYA2002-04094-C03. CB would like to acknowledge support from 
GENCAT (grant 2000FI 00376) and IEEC in Barcelona, and from grant GO3-4066X from SAO at Rutgers. KJB is supported by 
NASA grant NAG 5-7153.

\appendix

\section{A LIBRARY OF SYNTHETIC SPECTRA FOR THE ANALYSIS OF EJECTA EMISSION IN TYPE Ia SNRs}

In this Appendix, we introduce our library of synthetic spectra for the ejecta emission in Type Ia 
SNRs. The objective of this library is to provide observers with a complete set of synthetic spectra for the ejecta emission
in SNRs, calculated from an extensive grid of Type Ia SN explosion models, at different values of $t$, $\rho_{AM}$, and $\beta$. 
At present, our library includes more than 800 synthetic spectra in sequences like those presented in Figures \ref{fig-5} and 
\ref{fig-7} for models DDTe and B30U. For each model, we have generated synthetic spectra for several values in the ranges 
$430 \leq t \leq 5000$ yr;
$2 \cdot 10^{-25} \leq \rho_{AM} \leq 5 \cdot 10^{-24} \mathrm{g \cdot cm^{-3}}$; and $\beta_{min} \leq \beta \leq 0.1$.
In \S~\ref{subsec:Rationale}, we discuss these synthetic spectra in the context of the tools that are currently used for the 
analysis of ejecta emission in SNRs. In\S~\ref{subsec:Comparing} we comment on potential applications for our 
models. In \S~\ref{subsec:Caveats}, some important caveats and limitations of the models are discussed. 
Finally, in \S~\ref{subsec:Grid}, 
we introduce a number of Type Ia SN explosion models that, together with those presented in Paper I,
complete the exploration of the parameter space for thermonuclear supernovae. 

\subsection{Rationale} \label{subsec:Rationale}

The spectral analysis of the ejecta emission in young SNRs is a complex problem. Despite the spectacular increase
in the quality of the observations, it has proved very difficult to extract the relevant physical parameters from
these observations in a reliable way with the available tools. 
A frequent approach involves the fitting of several more or less sophisticated NEI components 
with varying abundances, electron temperatures, and ionization timescales to the observed X-ray spectra (several examples have been
cited in \S~\ref{subsec:Can3Ddoit?}). The results of applying this approach are not unique, and frequently very hard to interpret, 
because average parameters (like $T_{e}$ or $n_{e}t$) are assigned to a plasma whose physical properties have an enormous 
dynamic range, and where different chemical elements often emit under different conditions (see Figures \ref{fig-1}, \ref{fig-3},
and \ref{fig-4}). The determination of elemental 
abundances in the entire volume of shocked plasma, which is crucial for establishing the connection between the SNRs and the 
SN explosions that 
originated them, is particularly unreliable when it is based on this approach. Often, NEI models just provide estimates for 
the emission measures of the elements, under the assumption of a homogeneous composition, and the difference 
between the ratios of these emission measures and the true abundance ratios in the plasma can be several orders of 
magnitude (see section 4.2 of Paper I).

The synthetic spectra presented in Paper I and in the present work open new possibilities for the interpretation of X-ray 
observations of Type Ia SNRs. Without claiming to include all the complex physical processes at play in young SNRs (see 
\S~\ref{subsec:Caveats}), these synthetic spectra provide a much more accurate representation of the state of the
shocked ejecta in young Type Ia SNRs than the simple NEI models currently available within software packages like
XSPEC. Moreover, since the synthetic spectra are calculated from realistic SN explosion models, the connection between
the observed spectrum and quantities like the explosion energy or the amount of each element present in the ejecta are
easy to establish. The trade off is that the comparison between our synthetic spectra and X-ray observations is not necessarily 
a straightforward procedure.
 
\subsection{Comparing Models and Observations} \label{subsec:Comparing}

Several strategies with varying degrees of sophistication can be followed to compare our models to observations.
A somewhat crude possibility is to focus on derived quantities like $ \langle T_{e} \rangle$ and 
$ \langle \tau \rangle$, as we have done in \S~\ref{subsec:Can3Ddoit?}. While this can lead to interesting results, it is far better 
to perform more direct spectral comparisons using the library of synthetic spectra that we present here.
The most effective way to apply this library will depend on the specific observational constraints for the SNR
under study. In some cases, like the historical Galactic SNRs, the age will be known accurately, but the distance (and
hence the total integrated X-ray flux and the radius of the forward shock) will be more uncertain. In other cases, like the
SNRs in the Magellanic Clouds, the distance will be known, but the age will not. Reliable independent estimates
for $\rho_{AM}$ may or may not be available. In each case, there will be more than one way  to reduce the dimensionality of the 
problem. Rather than providing a recipe which may not be adequate for a specific case, we make here some suggestions which might 
prove useful in a more general context.

First, it is important to note that each synthetic spectrum is based on an underlying hydrodynamic model, so quantities such as the 
radius of the forward and the reverse shocks and their expansion parameters are available for each spectral model
(see Figure 3 and eqns. (3), (4), and (5) in Paper I). In principle, it is possible to reverse the problem, find out which 
hydrodynamic models agree better with the observations and thus reduce the number of synthetic spectra to consider. In doing so,
however, the limitations of 1D adiabatic hydrodynamics must be considered (see the following section). Second, the selection
of a particular synthetic spectrum from our library to represent the ejecta component in an observed X-ray spectrum may not be 
trivial. The substantial uncertainties in the atomic data and the relative simplicity of the models with respect to real SNRs 
will probably make it impossible to attain a statistically valid fit. Synthetic spectra like ours 
are more vulnerable to these factors, because there is little room for self-adjustment, in contrast to conventional NEI models
with variable abundances. If the emissivity of a particular line is underestimated in the spectral code used to generate 
our library, for instance, this cannot be compensated by artificially enhancing the abundance of that particular element, 
as in a conventional NEI model. Yet, even if some specific details of the observed spectrum cannot be reproduced, it is
often possible to find a model whose overall characteristics are in reasonable agreement with the observations.
Under these circumstances, a procedure needs to be devised in order to measure the degree of success of a specific 
synthetic spectrum. An example shall be provided in a forthcoming
paper on the Tycho SNR (Badenes et al., in preparation).

\subsection{Approximations and Caveats} \label{subsec:Caveats}

Our models are just a simplified representation of the complexity of young SNRs, and their limitations have 
to be considered when making comparison with observations. The crucial approximations were reviewed in sections 3.5 and 5 of Paper I, 
but it is important to revisit several issues here. 

The most important simplification is certainly the assumption of spherical symmetry. Any description of young SNRs in the framework 
of 1-D models is necessarily incomplete, because it does not include important processes such 
as ejecta clumping and dynamic instabilities at the contact discontinuity between shocked ejecta and shocked AM 
\citep{chevalier92:hydrod_instab_SNR,wang01:Ia-inst-clump}. The degree of ejecta clumping is crucial, and it is clear 
that our 1D models (and in particular,  the distribution of $\tau$ and $T_{e}$ for each element) would be invalidated if 
clumps with a large density contrast like those 
proposed by \citet{wang01:Ia-inst-clump} were to dominate the emission measure of the shocked ejecta in 
Type Ia SNRs. In this case, gross inconsistencies are expected to emerge from comparison of 1-D models with observations.
The degree of ejecta clumping strongly affects the morphology of the X-ray emission, and examination of this morphology 
in Type Ia SNRs should shed light on this issue. Multi-dimensional hydrodynamical simulations coupled with X-ray emission 
calculations could prove useful for this.

Another important issue whose impact on the X-ray spectra is hard to estimate is the effect of cosmic ray acceleration at the shocks. 
There is some indication that this process might affect the dynamics and X-ray spectra of the shocked AM without significantly 
modifying those of the shocked ejecta \citep{decourchelle00:cr-thermalxray}, but more detailed simulations are needed to shed light 
on this question \citep[see][for a discussion]{ellison05:cr+revshock}. 

To conclude this section, we comment on the importance of radiative losses, which have received some attention lately in the work of
Blinnikov et al. \citep[see][and references therein]{blinnikov04:lc+snr}. In Section 4 of \citet{hamilton84:ejecta}, it was 
shown that radiative losses always lead to catastrophic cooling in heavy element plasmas, driving the shocked material to infrared- 
and optically-emitting temperatures. Because no optically emitting knots with a composition dominated by heavy elements have been 
observed in Kepler, Tycho, or SN\,1006 \citep[see][]{blair91:kepler_optical,smith91:six_balmer_snrs}, we conclude that 
radiative losses are not dynamically important in young Type Ia SNRs under usual conditions. Radiative losses are not included
in our models in a self-consistent way, but we have extended the {\it a posteriori} monitoring of radiative losses described in 
section 3.5 of Paper I to the more unfavorable case of $\rho_{AM}=5\cdot10^{-24} \mathrm{g\cdot cm^{-3}}$. Our previous conclusions 
have been verified: radiative losses only affect the outermost layers of the models with the steepest ejecta density profiles. The 
values of $t_{rad}$, as defined in Paper I (the time when the calculated losses exceed 10\% of the specific internal energy in at 
least 5\% of the ejecta mass) for models with $t_{rad}<$5,000 yr are provided in Table \ref{tab-2}. In these models, our calculations 
for the properties (density, electron temperature, ionization state, and X-ray emission) of the layers that undergo radiative losses 
are not reliable close to or beyond $t_{rad}$, and some amount of infrared or optical emission should be expected from this region of 
the ejecta. The fact that such emission is not observed in Kepler, Tycho or SN\,1006 suggests that models which predict substantial
radiative losses in the ejecta are in conflict with observations of these historical SNRs.

\subsection{The Complete Grid of Type Ia SN Explosion Models} \label{subsec:Grid}
 
\subsubsection{One Dimensional Models}

Among the eight one dimensional Type Ia SN explosion models introduced in Paper I, one was a sub-Chandrasekhar explosion (SCH), 
one was a pure detonation (DET), two were pure deflagrations (DEFa and DEFf), two were delayed detonations (DDTa and DDTe), and two
were pulsating delayed detonations (PDDa and PDDe). These explosion paradigms, and the details involved in the calculation of the
models, are described in Paper I (Section 2 and Appendix). For the deflagrations, delayed detonations and pulsating delayed 
detonations, the models presented in Paper I were extreme cases, obtained by considering the highest and lowest reasonable values 
of the parameters involved in each calculation. In the case of the deflagration models, the relevant parameter is $\kappa$, which 
controls the propagation velocity of the subsonic flame. For the delayed detonation and pulsating delayed detonation, the parameters 
are $\rho_{tr}$, which determines the density at which the transition from deflagration to detonation occurs, and $\iota$, which 
determines the flame velocity in the deflagration stage. All these parameters are defined in the Appendix of Paper I.
By varying these parameters, we have generated four more deflagrations 
(DEFb, DEFc, DEFd, and DEFe), four more delayed detonations (DDTb, DDTbb, DDTc, and DDTd), and three more pulsating delayed 
detonations (PDDb, PDDc, and PDDd). In Table \ref{tab-3} and Figure \ref{fig-8}, we present the nucleosynthetic output, chemical 
composition profiles and density profiles of these intermediate models that complete the exploration of the parameter space. 

\subsubsection{Three Dimensional Models}

\begin{figure}

  \centering
  \includegraphics[scale=0.55]{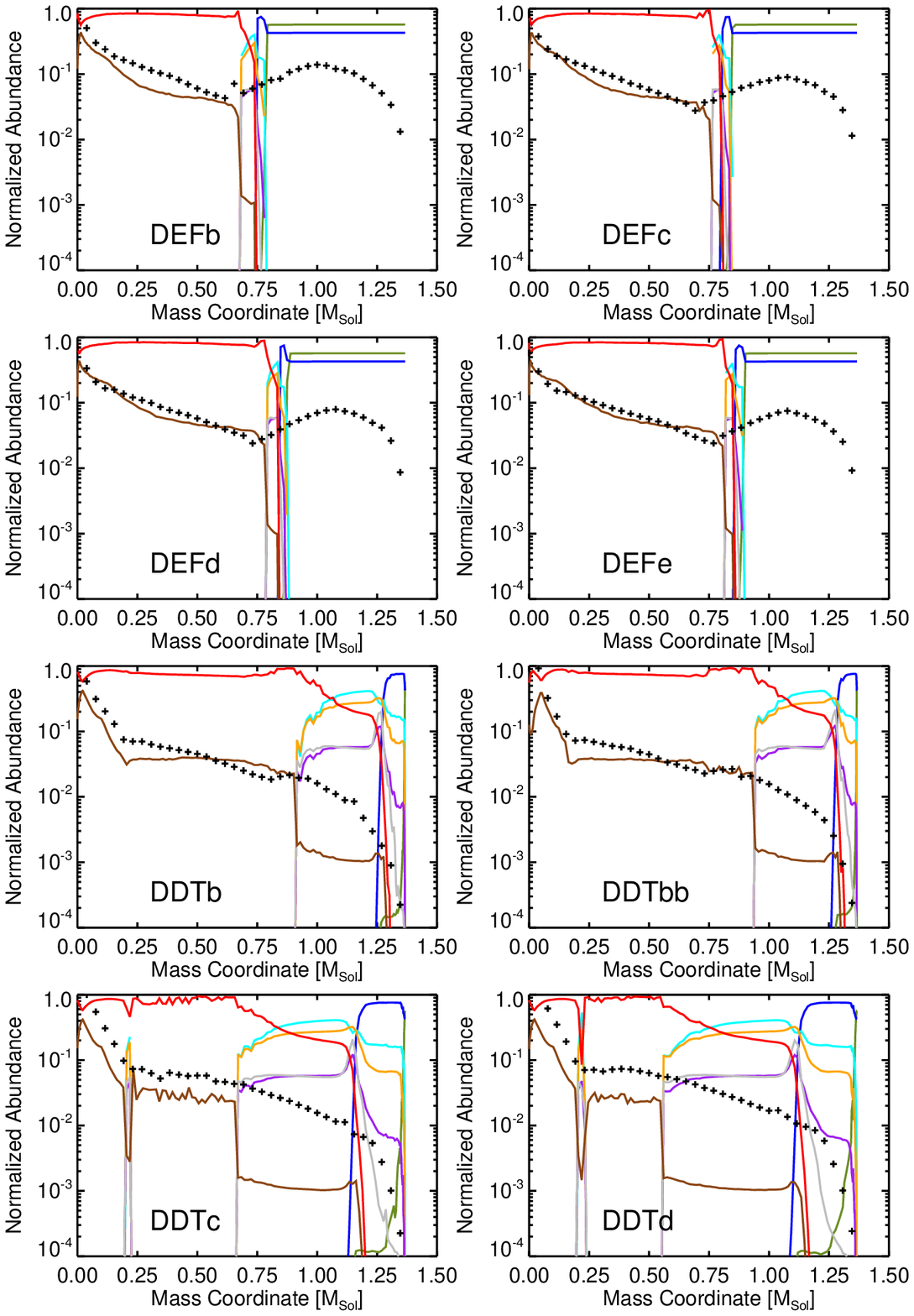}
  \includegraphics[scale=0.55]{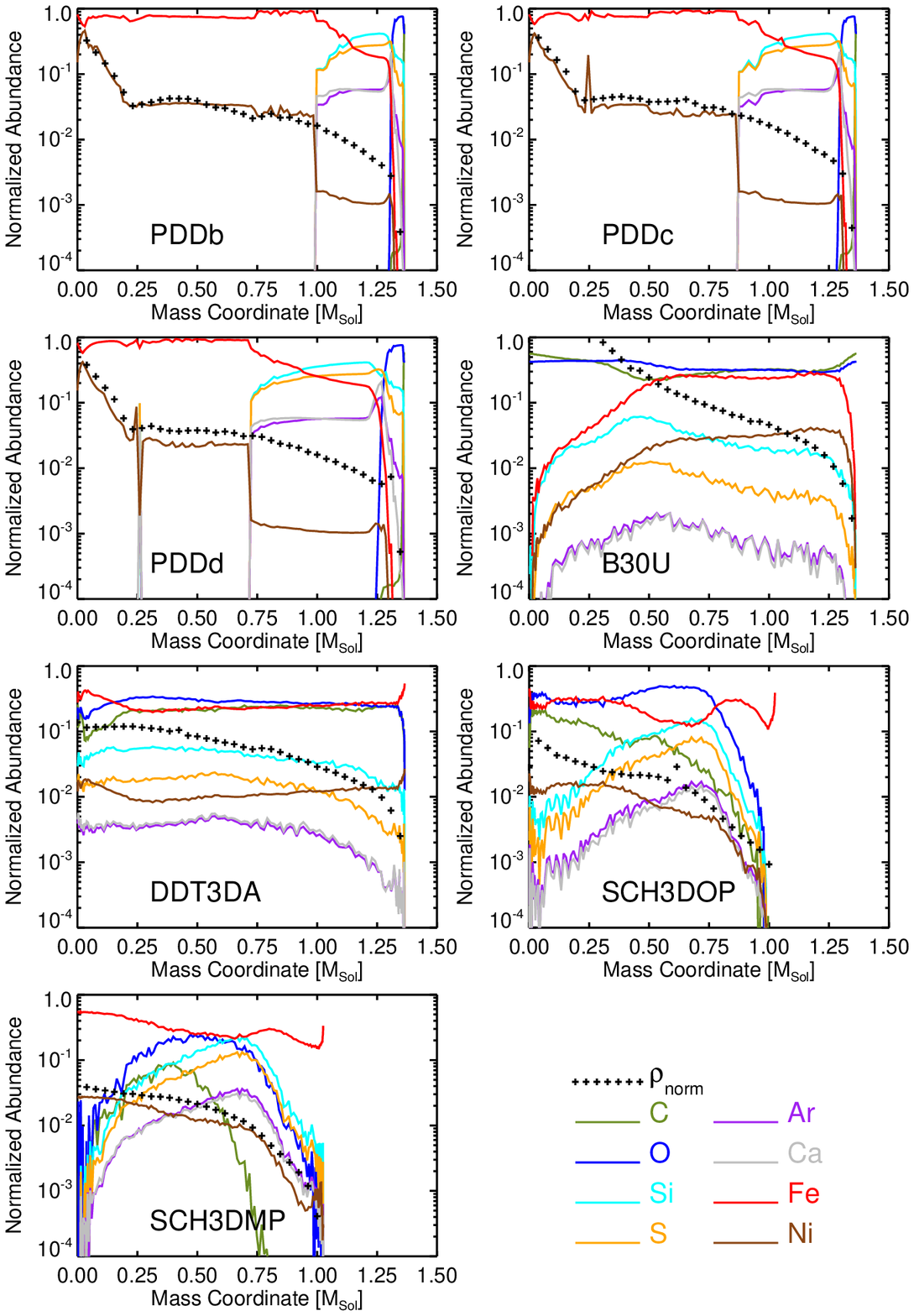}

  \caption{
   Chemical composition and density profiles for the Type Ia SN explosion models that were not presented in Paper I.
   The abundances represented here are number abundances after the decay of all short lifetime isotopes. The density 
   profiles ($\rho_{\mathrm norm}$) are represented at $t=10^{6}$ s after the SN explosion, and they have been
   normalized by $\rho_{n}=10^{-11}\, {\mathrm g \cdot cm^{-3}}$. See the on-line edition for a color
   version of this Figure.\label{fig-8}}

\end{figure}

A brief discussion on the state of the art in 3D calculations of thermonuclear SN explosions can be found in \S~\ref{subsec:3Dmodels}
of this work; for a review see \citet{bravo05:SNIa_3D_review}. Without going into the details of how these 
3D models are calculated, here we present four one-dimensional mappings of 3D models that are representative of the current trends. 
Their main characteristics are given in Figure \ref{fig-8} and Table \ref{tab-4}. Model B30U is a 3D deflagration from 
\citet{garcia-senz03:Ia3D}, very similar 
to the models by \citet{gamezo03:Ia3D} and \citet{travaglio04:3D} (see discussion in \S~\ref{sec:Expanding}). Model DDT3DA is a 
3D version of the delayed detonation paradigm \citep{garcia-senz03:DDT3D}. 
In this model, a detonation was artificially inducted in those regions were the flame resulting from the turbulent deflagration
phase was well described by a fractal surface of dimension larger than 2.5. We note that this particular model also results in 
very well mixed ejecta, and in fact the properties of its X-ray emission in the SNR phase are very similar to 
those of model B30U. 
Other delayed detonations in 3D calculated with different assumptions for the induction of the detonation
result in more stratified ejecta \citep{gamezo04:DDT3D,gamezo05:DDT3D}. For a comparison between these two kinds of three dimensional 
delayed detonations, see Table 1 and the accompanying text in \citet{bravo05:SNIa_3D_review}. Finally, two 3D sub-Chandrasekhar
models from \citet{garcia-senz99:SCH3D} have also been included in the grid. Model SCH3DOP is a sub-Chandrasekhar explosion
calculated in 3D where the layer of degenerate He was ignited at one single point, while in SCH3DMP the ignition happened
at five different points. It is worth noting that none of the 3D models has been followed for a sufficient time to account 
for the effects of the decay of $^{56}$Ni on the density profiles.

\clearpage
\bibliographystyle{apj}


\clearpage 

\clearpage
\centering
\setlength{\tabcolsep}{0.02in}
\begin{deluxetable}{ccccccccc}
  \tabletypesize{\scriptsize}
  \tablecaption{Spectral properties of Fe and Si in Type Ia SNRs\label{tab-1}}
  \tablecolumns{9}
  \rotate
  \tablehead{
    \colhead{SNR} &
    \colhead{Reference} &
    \colhead{Age} &
    \multicolumn{3}{c}{Si component \tablenotemark{a}}  &
    \multicolumn{3}{c}{Fe component} \\
    \colhead{} &
    \colhead{} &
    \colhead{} &
    \colhead{Spectral model} &
    \colhead{$kT$} &
    \colhead{$log\,n_{e}t$} &
    \colhead{Spectral model} &
    \colhead{$kT$} &
    \colhead{$log\,n_{e}t$} \\
    \colhead{} &
    \colhead{} &
    \colhead{[yr]} &
    \colhead{} &
    \colhead{[keV]} &
    \colhead{[$\mathrm{cm^{3} s}$]} &
    \colhead{} &
    \colhead{[keV]} &
    \colhead{[$\mathrm{cm^{3} s}$]}
  }
  \startdata
  Tycho & \citet{hwang98:tycho-Feemission} & 432 & NEI, single $T_{e}$, & 
  $0.86$ & $\backsim 11$ & NEI, single $T_{e}$, & $> 1.7$ & $\sim 9$ \\
  & & & single $n_{e}t$ & & & single $n_{e}t$ & & \\
  Kepler\tablenotemark{b} & \citet{kinugasa99:kepler-ASCA} & 404 & NEI, single $T_{e}$, &
  $0.77\pm0.06$ & $10.42^{+0.06}_{-0.05}$ &  NEI, single $T_{e}$, & $>6$ & $9.53^{+0.07}_{-0.01}$ \\ 
  & & & single $n_{e}t$ \tablenotemark{c} & & & single $n_{e}t$ \tablenotemark{c} & & \\
  0509-67.5 & \citet{warren03:0509-67.5} & $< 1,000$ & Plane-parallel &
  $3.13\pm0.55$ & $9.93\pm0.02$ & Plane-parallel & $10.0^{+\infty}_{-5.44}$ & $10.53\pm0.02$ \\
  & & & NEI shock \tablenotemark{d} & & & NEI shock \tablenotemark{d} & & \\
  N103B & \citet{lewis03:N103B} & $< 2,000$ \tablenotemark{e} & Plane-parallel &
  $\backsim 1$ & $>12$ & Plane-parallel & $> 2$ & $\sim 10.8$ \\
  & & & NEI shock & & & NEI shock & & \\
  \enddata

  \tablenotetext{a}{In all the cases listed here, the 'Si component' also included some Fe, as well as other elements.}
  \tablenotetext{b}{The identification of this SNR as Type Ia is controversial, see \citet{blair04:kepler}.}
  \tablenotetext{c}{\citet{kinugasa99:kepler-ASCA} also fitted the spectrum with an NEI component for the Fe K$\alpha$ line
    plus a more sophisticated model for the rest of the shocked ejecta and AM, based on the self-similar solutions of 
    \citet{chevalier82:selfsimilar}, coupled to NEI calculations in a plasma with homogeneous abundances. However,
    they do not give explicit values for $kT$ and $n_{e}t$ of the reverse shock in this model. For simplicity, we use the
    results of their two-component NEI model, which gives a fit of similar quality.}
  \tablenotetext{d}{We list the results for the best fit model in \citet{warren03:0509-67.5}, which includes a nonthermal
    continuum (their model S). Assuming a thermal continuum (their model H), these authors obtain similar results: 
    $kT_{Si}=2.23\pm0.29$, $log\,n_{e}t_{Si}=9.94\pm0.02$, $kT_{Fe}=10.00^{+\infty}_{-4.14}$, $log\,n_{e}t_{Fe}=10.53\pm0.02$.}
  \tablenotetext{e}{\citet{hughes95:typing_SN_from_SNR}.}

\end{deluxetable}

\clearpage
\begin{deluxetable}{ccc}
  \tablecaption{Values of $t_{rad}$ \tablenotemark{a} \label{tab-2}}
  \tablecolumns{3}
  \tablehead{
    \colhead{Model} &
    \colhead{$\rho_{AM}=10^{-24}\mathrm{g\cdot cm^{-3}}$} &
    \colhead{$\rho_{AM}=5\cdot10^{-24}\mathrm{g\cdot cm^{-3}}$}
  }
  \startdata
  DEFa & $3.0\cdot10^{10}$ s & $2\cdot10^{10}$ s \\
  DEFc & $2.7\cdot10^{10}$ s & $1.6\cdot10^{10}$ s \\ 
  DEFf & $2.4\cdot10^{10}$ s & $1.2\cdot10^{10}$ s \\
  PDDe & - & $1.5\cdot10^{11}$ s \\
  \enddata

  \tablenotetext{a}{Only models with $t_{rad}<$5,000 yr ($1.58\cdot10^{11}$ s) are listed. The values of  $t_{rad}$ 
    for other DEF models (DEFb, DEFd and DEFe) are comparable. All calculations were done with $\beta=\beta_{min}$}

\end{deluxetable}

\clearpage
\setlength{\tabcolsep}{0.07in}
\begin{deluxetable}{ccccccccccccc}
  \tabletypesize{\scriptsize}
  \tablecolumns{13}
  \tablecaption{Properties of the additional Type Ia 1D explosion models\label{tab-3}}
  \tablehead{
    \colhead{Model} &
    \colhead{Para-} &
    \colhead{$\rho_{tr}$} &
    \colhead{$M_{\mathrm ejecta}$} & 
    \colhead{$E_{k}$} & 
    \colhead{${\mathcal M}_{\mathrm max}$ \tablenotemark{b}} & 
    \colhead{$\Delta {\mathcal M}_{\mathrm 15}$ \tablenotemark{b}} &   
    \colhead{$M_{\mathrm{Fe}}$} &
    \colhead{$M_{\mathrm{C+O}}$} &
    \colhead{$M_{\mathrm{Si}}$} &
    \colhead{$M_{\mathrm{S}}$} &
    \colhead{$M_{\mathrm{Ar}}$} &
    \colhead{$M_{\mathrm{Ca}}$} \\  
    \colhead{} &
    \colhead{meter \tablenotemark{a}} &
    \colhead{[${\mathrm g \cdot cm^{-3}}$]} &
    \colhead{[${\mathrm M_{\sun}}$]} &
    \colhead{[$10^{51}$ erg]} &
    \colhead{[mag]} &
    \colhead{[mag]} &
    \colhead{[${\mathrm M}_{\sun}$]} &
    \colhead{[${\mathrm M}_{\sun}$]} &
    \colhead{[${\mathrm M}_{\sun}$]} &
    \colhead{[${\mathrm M}_{\sun}$]} &
    \colhead{[${\mathrm M}_{\sun}$]} &
    \colhead{[${\mathrm M}_{\sun}$]}
  }
  \startdata
  DEFb & 0.08 & & 1.37 &
  0.64 & -19.14 & 0.94 & 0.61 &
  0.61 & 0.025 & 0.017 & 0.0040 &
  0.0043 \\
  DEFc & 0.10 & & 1.37 &
  0.74 & -19.29 & 0.99 & 0.68 &
  0.55 & 0.021 & 0.014 & 0.0032 &
  0.0032 \\
  DEFd & 0.12 & & 1.37 &
  0.80 & -19.34 & 1.02 & 0.71 & 
  0.52 & 0.021 & 0.014 & 0.0032 &
  0.0034 \\
  DEFe & 0.14 & & 1.37 &
  0.81 & -19.29 & 0.98 & 0.73 &
  0.49 & 0.021 & 0.013 & 0.0029 &
  0.0028 \\
  \hline 
  DDTb & 0.03 & $2.6\cdot10^{7}$ & 1.37 &
  1.36 & -19.67 & 1.11 & 0.98 &
  0.05 & 0.10 & 0.084 & 0.022 &
  0.027 \\
  DDTbb & 0.01 & $2.5\cdot10^{7}$ & 1.37 &
  1.31 & -19.66 & 1.12 & 0.99 &
  0.05 & 0.10 & 0.084 & 0.022 &
  0.027 \\
  DDTc & 0.03 & $2.2\cdot10^{7}$ & 1.37 &
  1.16 & -19.51 & 1.11 & 0.80 &
  0.12 & 0.17 & 0.13 & 0.033 &
  0.038 \\
  DDTd & 0.03 & $1.5\cdot10^{7}$ & 1.37 &
  1.08 & -19.30 & 0.94 & 0.72 &
  0.14 & 0.20 & 0.15 & 0.037 &
  0.043 \\
  \hline 
  PDDb & 0.03 & $2.2\cdot10^{7}$ & 1.37 &
  1.36 & -19.72 & 1.14 & 1.04 &
  0.03 & 0.085 & 0.070 & 0.018 &
  0.022 \\
  PDDc & 0.03 & $1.5\cdot10^{7}$ & 1.37 &
  1.25 & -19.64 & 1.11 & 0.98 &
  0.04 & 0.11 & 0.093 & 0.024 &
  0.029 \\
  PDDd & 0.03 & $1.2\cdot10^{7}$ & 1.37 &
  1.24 & -19.53 & 1.04 & 0.89 &
  0.05 & 0.15 & 0.13 & 0.034 &
  0.041 \\
  \enddata

  \tablenotetext{a}{The parameter given is $\kappa$ for the DEF models and
    $\iota$ for the DDT and PDD models (see the Appendix of Paper I for details).} 
  \tablenotetext{b}{The values of ${\mathcal M}_{\mathrm max}$ and $\Delta {\mathcal M}_{\mathrm 15}$ for the
    light curves were calculated by I. Dom\'inguez (private communication, 2003).}
\end{deluxetable}

\clearpage
\begin{deluxetable}{ccccccccc}
  \tabletypesize{\scriptsize}
  \tablecolumns{9}
  \tablecaption{Properties of the 3D Type Ia explosion models\label{tab-4}}
  \tablehead{
    \colhead{Model} &
    \colhead{$M_{\mathrm ejecta}$} & 
    \colhead{$E_{k}$} &    
    \colhead{$M_{\mathrm{Fe}}$} &
    \colhead{$M_{\mathrm{C+O}}$} &
    \colhead{$M_{\mathrm{Si}}$} &
    \colhead{$M_{\mathrm{S}}$} &
    \colhead{$M_{\mathrm{Ar}}$} &
    \colhead{$M_{\mathrm{Ca}}$} \\  
    \colhead{} &
    \colhead{[${\mathrm M_{\sun}}$]} &
    \colhead{[$10^{51}$ erg]} &
    \colhead{[${\mathrm M}_{\sun}$]} &
    \colhead{[${\mathrm M}_{\sun}$]} &
    \colhead{[${\mathrm M}_{\sun}$]} &
    \colhead{[${\mathrm M}_{\sun}$]} &
    \colhead{[${\mathrm M}_{\sun}$]} &
    \colhead{[${\mathrm M}_{\sun}$]}
  }
  \startdata
  B30U & 1.37 & 0.42 & 0.53 &
  0.66 & 0.045 & 0.011 & 0.0019 &
  0.0017 \\
  DDT3Da & 1.37 & 0.78 & 0.76 &
  0.38 & 0.063 & 0.027 & 0.0066 &
  0.0072 \\
  SCH3DOP & 1.02 & 1.14 & 0.58 &
  0.23 & 0.064 & 0.035 & 0.0093 &
  0.0077 \\
  SCH3DMP & 1.02 & 1.19 & 0.67 &
  0.07 & 0.081 & 0.054 & 0.019 &
  0.017 \\
  \enddata

\end{deluxetable}

\end{document}